\documentstyle[11pt,epsf]{article}

\input psfig
\def\beq{\begin{eqnarray}}
\def\eeq{\end{eqnarray}}
\def\bea{\begin{eqnarray*}}
\def\eea{\end{eqnarray*}}

\def\centeron#1#2{{\setbox0=\hbox{#1}\setbox1=\hbox{#2}\ifdim
\wd1>\wd0\kern.5\wd1\kern-.5\wd0\fi
\copy0\kern-.5\wd0\kern-.5\wd1\copy1\ifdim\wd0>\wd1
\kern.5\wd0\kern-.5\wd1\fi}}
\def\ltap{\;\centeron{\raise.35ex\hbox{$<$}}{\lower.65ex\hbox{$\sim$}}\;}
\def\gtap{\;\centeron{\raise.35ex\hbox{$>$}}{\lower.65ex\hbox{$\sim$}}\;}

\def\singleandthirdspaced{\baselineskip=\normalbaselineskip\multiply
    \baselineskip by 130\divide\baselineskip by 100}
\def\singlespaced{\baselineskip=\normalbaselineskip}

\newcommand{\newc}{\newcommand}
\newc{\qbar}{{\overline q}}
\newc{\Kahler}{K\"ahler }
\newc{\deltaGS}{\delta_{\rm GS}}
\begin{document}
\begin{titlepage}
\begin{flushright}
{\large hep-th/0008058 \\ SCIPP-00/26\\
}
\end{flushright}

\vskip 1.2cm

\begin{center}

{\LARGE\bf Some Issues in Flat Direction Baryogenesis}

\vskip 1.4cm

{\large  Alexey Anisimov and Michael Dine}
\\
\vskip 0.4cm
{\it Santa Cruz Institute for Particle Physics,
     Santa Cruz CA 95064  } \\

\vskip 4pt

\vskip 1.5cm

\begin{abstract}
Motivated by recent developments, we explore some issues in
Affleck-Dine baryogenesis.  We consider in greater detail the role
of thermal effects in the production of baryon number.  We find
that these effects are important even for rather flat
potentials, and obtain somewhat different estimates
of the baryon asymmetry than those in the literature.
We also consider the decay of the condensate, and possible
implications of these observations for the formation of Q-balls.
\end{abstract}

\end{center}

\vskip 1.0 cm

\end{titlepage}
\setcounter{footnote}{0} \setcounter{page}{2}
\setcounter{section}{0} \setcounter{subsection}{0}
\setcounter{subsubsection}{0}

\singleandthirdspaced


\section{Introduction}

Through the years, there have been a number of proposals to understand
the origin of the
asymmetry between matter and antimatter.  If nature is
supersymmetric, baryogenesis through the semiclassical evolution
of a scalar condensate provides a mechanism which is particularly
attractive,
both for simplicity and efficiency.  In the
original analysis, it was supposed that in a theory like the MSSM
there would be some exact flat directions, in the limit of
unbroken supersymmetry, and that it would be natural for fields in
these directions to start off with very large expectation values.
Several sources of baryon number and CP violation could be
imagined, which could easily generate an enormous baryon
asymmetry\cite{affleckdine}.

However, the assumptions of the original work were rather naive,
and in considering Affleck-Dine baryogenesis, there are a number of
issues one must address\cite{drt}.
\begin{itemize}
\item
Just how flat are the flat
directions?  Invariably, higher dimension terms in the
superpotential lift the flat directions, unless they are protected
by discrete R symmetries (or perhaps by stringy effects).  For
most $D$-flat directions, there are candidate terms in the
superpotential quartic in the fields (or even low order).
\item
In considering the evolution of the flat directions, one must take
into account the supersymmetry breaking effects of the early
universe.  Not only are these effects dominant in the period when
$H > m_{3/2}$, they can provide an explanation why the fields
start at large values.  In general, the effective masses of
scalars are of order $H$.  If these masses are negative,
they force the fields to large values, providing the initial
conditions for baryogenesis.  In the case of flat directions
lifted by quartic terms in the superpotential,
$\delta W = {\phi^4 \over M}$
one has $\phi^2 \sim H M$.  More generally, one has, for a
superpotential
\beq
W_n = {\phi^{n+3} \over M^n},
~~~~~~
{\phi^{2n+2} \sim H^2 M^{2n}}.
\eeq
The evolution of the system, as well as the ultimate amount of
baryon production, are crucially dependent on $n$.
\item
In considering how baryon number is produced, one must determine
the sources of baryon number violation and CP violation.  In most cases
the
most important effect is supersymmetry violating $A$ terms.  In
addition to terms proportional to $m_{3/2}$, there
are also terms scaled by $H$, and it is usually the mismatch
in the phases of the various terms which accounts for baryon
production.  For example, if one has a superpotential term
proportional to $\phi^{n+3}$, as above, one expects an $A$ term
proportional to $H(\phi^{n+3} + {\rm c.c.})$.  This term typically
violates both baryon (and/or lepton) number and CP.  In generating
a baryon number, it is crucial that the phase of this term is
different from the phase that exists in flat space.
\item
One must carefully consider the mechanisms by which the condensate
may decay.    In \cite{drt}, it was pointed out that the condensate
generally decays as a result of scattering with the thermalized
decay products of the inflaton.  These have a temperature
which behaves roughly as
\beq
T \sim (T_R^2 H M_p)^{1/4},
\label{temperature}
\eeq
and this is substantially higher than the
reheat temperature for some period.   The fields which couple to
$\phi$, the AD scalar, gain mass.  $\phi$, however, damps as
$1/t$, so eventually some of the fields have mass of order $T$
and come to equilibrium.  The scattering of these particles off
the condensate causes the condensate to quickly evaporate.
Generally, these interactions preserve $B$ and $L$, so the
previously produced baryon number is unchanged.
\item
Concerning the last point, it was argued in \cite{drt}
that that if the
condensate is not very large, thermal effects can even lead to
evaporation of the condensate before the baryon number is
produced.  If $y$ is the smallest
Yukawa coupling in a particular flat direction, there are
fields with mass of order $y \phi$.  If $y$ is of order
$10^{-2}-10^{-4}$, and if $\phi$ is not too large,
then it is possible for some of these fields to be in thermal
equilibrium.
Using eqn. \ref{temperature}, if
$T_R = 10^{8}$ GeV,
for example, then at $H \approx m_{3/2}$, $T \approx  10^{9}$ GeV;
if the reheating temperature is $10^{10}$, then the temperature is
of order $10^{10}$ at this time.
On the other hand, in the case $n=1$,
the initial amplitude of
the $\phi$ oscillations (when $H \sim m_{3/2}$) are of order
$ \phi \sim \sqrt{m_{3/2} M_p} \sim 10^{11}$ GeV, so if
some Yukawa couplings are smaller than $10^{-3}$ or so,
then even if the reheating temperature is of order
$10^8$ GeV,  such states will be in thermal
equilibrium. Even though the Yukawa coupling is small,
the rate for scattering of particles off the condensate is
enormous compared to the age of the universe in these circumstances.
Assuming the mass of the condensate particles is of order
$m_{3/2}$, a typical cross section for scattering off
the condensate will be of order
\beq
\sigma \sim {y^2 \alpha \over m_{3/2}T}
\eeq
Multiplying by $T^3$, yields a reaction rate compared to the
Hubble density  of order
\beq
{\Gamma \over H} \sim y^2 \alpha {M_p \over m_{3/2}}.
\eeq
In other words, the condensate disappears almost immediately.
\end{itemize}

As a result of these latter considerations, it was argued in
\cite{drt} that for $n=1$, in order to produce a baryon asymmetry,
one must either consider flat directions with only large Yukawa
couplings, or one must assume a low reheat temperature, in order
to have baryogenesis.

Recently,
Campbell et al \cite{campbell} have made an important observation
concerning this picture.   They note that if the
scalar fields are not too large, so that some of the fields coupled to the
condensate are in thermal equilibrium, not only is there rapid
evaporation, but there are also large, positive, thermal masses for the
scalars, as a result of which they oscillate earlier than
otherwise expected.  Moreover, the evaporation of the condensate
is slow enough that the
system undergoes several oscillations before it disappears.
Finally, they argue that there are similar thermal $A$
terms, as a result of which one can produce a significant
asymmetry before the condensate evaporates.

We will explore these issues more thoroughly here.  These thermal
effects are particularly important in the case $n=1$, and can
be important for $n=2$.  We will see that generically the thermal $A$ terms
exploited by \cite{campbell} to generate
the required CP violation vanish.  We study other possible sources of
CP violation, and find that in some circumstances these
can be effective.

We will see (as observed in \cite{campbell}), if thermal
masses are important, the estimate above of the
rate of evaporation is far too large.  Clearly, in the cases
where particles coupled to the condensate are in thermal
equilibrium, the appropriate
mass for the condensate particles is not the zero temperature mass
but the effective thermal mass (this, after all, determines the
oscillation rate).  This means that the evaporation cross section
is far smaller than in the earlier estimates, and in fact the
condensate evaporates well after the formation of the asymmetry.
We do not agree in detail with the estimate of the cross section
in \cite{campbell}, but a reliable calculation requires the use of
real time finite temperature methods, and we will only
make crude estimates here.  We will assume that the
cross section is dominated by processes such as scattering of
thermalized fermions off the condensate.  This cross section is
proportional to two powers of the Yukawa coupling, $y^2$ and of
the gauge coupling, $g^2$, divided by the center of mass energy.
This energy is of order $\sqrt{y T^2}$, the factor $yT$
corresponding to the thermal mass of the condensate particles, and
the factor of $T$ the typical energy of the fermions.  In other
words, we will use as our working formula for the reaction rate:
\beq
\Gamma= {y \alpha T}.
\eeq

Once one has considered thermal masses, one realizes that
even in the case where the fields coupled to the condensate are
not in thermal equilibrium there are other thermal effects which
are important at high temperatures.  Couplings to light fields
through higher dimension operators leads to contributions to the
condensate potential which can dominate for a significant period, and
which are potentially crucial to determining the baryon number,
even for rather large values of $n$ ($n=2-4$, and possibly larger
depending on parameters).

\section{Thermal Effects}

To get some feeling for the issues involved, suppose, for
definiteness, that the inflaton, $I$, has a mass of order $10^{13}$ GeV,
while the inflaton amplitude immediately after inflation is of
order $M_p$, and its associated auxiliary
field, $F_I \sim 10^{13} M_p$.  Then the inflaton width is of order $10^3$ GeV or
so, while the Hubble constant during inflation is
of order $10^{13}$.   After inflation, the temperature quickly rises to
order $10^{13}$ GeV, and then falls roughly as $t^{-1/4}$.  The
inflaton decays when $H \sim 10^{3}$ GeV, with a reheat temperature of
order $10^{10}$ GeV.  This reheat temperature is perhaps somewhat high
from the point of view of gravitino production, but we view it as
a conservative choice from the perspective of the issues we
address in this paper.

If the flat directions are lifted
by a term with $n=1$ (i.e. a $\phi^4 \over M_p$ term in the
superpotential),
then when inflation ends,
$\phi \approx \sqrt{H M} \sim 10^{16}$.
If there are Yukawa couplings (generically
denoted $y$) in the flat direction (as there
typically are) less than about $10^{-3}$, then the corresponding
fields are in thermal equilibrium.  For larger Yukawa couplings,
the system still may come to thermal equilibrium well before the
inflaton decays.  As a result, the scalar fields
have thermal masses of order $yT$.  Assuming a negative contribution
to the masses of order $H^2$, theses lead to oscillations
once $yT \approx H$.  This is long before
$H \sim m_{3/2}$.  Thermal effects also lead to evaporation.  Typically, as pointed
out in \cite{campbell}, however,
the evaporation timescale is somewhat suppressed relative to the
oscillation
timescale, and one might hope to produce baryons.  Indeed, examining
our
formula for the scattering rate, we see that this differs from the
oscillation time by a factor of $\alpha$.

On the other hand, in the picture developed in \cite{drt}, the
crucial element in the generation of the baryon asymmetry is
a misalignment between the phases of the $A$ terms due to the
oscillating inflaton and
those which exist at zero curvature.  When $\phi$ begins to
oscillate at times of order $m_{3/2}$, the difference in these
phases leads to a net torque.  If the fields start to oscillate
much earlier, however, the effect of the zero curvature terms is
suppressed, and one may have difficulty generating an asymmetry.
(One can see that misalignment is necessary by noting that
otherwise one can eliminate any phase by a suitable field redefinition.)

The authors of \cite{campbell} argued that there are
additional $A$ terms, proportional to $T$, and that these could be
responsible for the asymmetry.  It is easy to see, however, that
this is not the case in a generic situation.
The issue is one of symmetries.  The usual
picture for the formation of the $A$ terms is to suppose that
there is an inflaton coupling of the form
\beq
\int d^4 \theta {I^* \over M_p} f(\phi) + c.c.
\label{aterms}
\eeq
which gives an $A$ term scaled by $H$.  It is important
here that the coupling to the dilaton breaks the $R$
symmetry of the renormalizable (susy-preserving) terms of the MSSM.
The authors of \cite{campbell} consider terms in the
superpotential of the form (taking the case $n=1$ for simplicity)
\beq
W = {\phi^{4} \over M} + h \phi \chi \chi
\eeq
where $\chi$ represents a field coupled to $\phi$ with
a sufficiently small Yukawa coupling, $h$, that it is in thermal
equilibrium.  Then the potential includes terms such as
\beq
\delta V = {\phi^3 \over M} \chi^* \chi^*.
\label{campbellw}
\eeq
They then assumed that in the thermal bath, $<\chi \chi> \sim
T^2$, so that one has, effectively, a quite large $A$ term.
However, in general, symmetries suppress this correlation
function, and the result is quite a bit smaller.  In particular,
the superpotential of eqn. \ref{campbellw} respects an $R$
symmetry, under which the $\chi$ fields transform, so we must ask
what terms violate this symmetry.  The $A$ terms, such as $H
{\phi \chi \chi}$ are examples of such terms, as are the gaugino
mass terms. Using this coupling, one has:
\beq
\langle \chi \chi \rangle = hHT\phi \int {d^3 k \over (2 \pi)^3}{1 \over
(k^2 +
m_{\chi}^2)^2}
\eeq
$$~~~~~~\approx {hHT \phi \over 2 \pi m_{\chi}}$$
where $m_{\chi}$, by assumption, is of order $T$
(in general,  (e.g. if $\chi$ couples
to gauge bosons of an unbroken symmetry, then the mass is of order $g^2
T^2$).
Even if one supposes that $m_{\chi}$ is smaller, e.g. $m_{\chi} \sim H$,
the resulting $A$ term is of order
\beq
V_A \approx h^2 {T \over M_p} \phi^4
\eeq
which, for $h \sim 10^{-2}$, is smaller than the non-thermal
term.
It is possible that one could find larger sources of symmetry
violation in particular models, but we believe that this is the
generic behavior; there is no temperature enhancement of the
$A$ term.

There are other possible sources of $A$ terms which can be
relevant.
The largest
contribution is likely to come from terms which behave like
$H {I \over M_P}$, where $I$ is the inflaton term.  These terms
are not drastically suppressed at the time when oscillations
begin, and they have a different time dependence than the leading
$A$ term,
which is simply proportional to $H$.  We will see in the next
section that these terms, in the case $n=1$, may barely yield an adequate
asymmetry.  In the case $n=2$, things are better.

These additional $A$ terms arise in a simple way.  In addition to
the term of eqn. \ref{aterms}, there may
be a coupling of the form:
\beq
\delta W = {1 \over M_p}(a I + b {I^2\over M_p})\phi^n,
\label{isquared}
\eeq
where $a$ and $b$ are complex constants.
$I$ decreases as $t^{-1}$, so the second term is suppressed by
${I_o\over M_p}{t_o \over t}$, and this need not be a very
large suppression.
These two terms need not have the same phase, so there is
the possibility of generating a reasonable baryon number.
We will explore the effects of these terms shortly.

Another possibility, which we will not explore in great detail here,
arises in the higher $n$ cases.  If the suppression of, say, the
$n=1$ terms arises because of discrete symmetries, it is possible
that there is less suppression of the $A$ terms with smaller $n$.
To see this, let us consider the
structure of eqn.[\ref{aterms}] in more detail.  As an example, take the
flat direction labeled by $H_u L$.  In order that this direction be
lifted by terms of order, say $(H_u L)^3/M_p^2$ (corresponding to $n=2$),
one might hypothesize
a suitable discrete symmetry.  Now $A$ terms are generated by terms
involving the inflaton field.  By simply postulating a  suitable $R$
transformation for $I$, one might allow the coupling $I (H_U
L)^2$.  In this case, the effective $A$ term is much larger than
naively expected; it is not of order $H$, but of order $H
(M^2/(H_U L))$.  Indeed, these terms can be so large as to require
modification of the whole picture.
One might try to do the same thing in the case $n=2$, but the
presence of a large mixing term, $m_{3/2} \mu H_U L$
might lead to phenomenological difficulties.


All of this is relevant only to directions which are not too flat.
Otherwise, the AD field is very large during the relevant time
period, and the fields to which it couples are very massive, with
mass much larger than the temperature.  This is the case
for $n \ge 2$ or so (the
precise value depending on the value of the Yukawa coupling).
In these other cases, however, there are other thermal effects
which much be considered.  In particular, integrating out massive
fields can generate couplings of the AD field to remaining light
fields, such as
\beq
{A \over 16 \pi^2} \ln(\vert \phi \vert^2)F_{\mu \nu}^2.
\eeq
To understand the effect of these terms at finite temperature,
note that this is a correction to the associated gauge coupling.
So we can compute the finite temperature $\phi$ potential by
simply calculating the free energy of the gauge theory as a
function of the gauge coupling.

As an example, consider the flat direction $H_U L$.  In this flat
direction, the unbroken gauge group is $SU(3) \times U(1)$.  The
$u$ quarks gain mass in this direction, as do the right handed leptons and
$H_D$. The coefficient, $A$, is obtained by integrating out these
fields:  $A= 3$.  To understand the effect of this coupling, note that
for slowly
varying $\phi$, this is just a modification of the $SU(3)$
coupling.  So the leading effect of this term can be determined by
considering the free energy as a function of the coupling
constant.  The leading contribution of gluons, gluinos and
quarks to the free energy can be readily obtained from
calculations in the literature:
\beq
\delta \Omega = {N_g \over 144} (5N_f/4 + 7N/2) g^2 T^4
\eeq
The contribution of the scalars requires an additional computation
which we have not found in the literature, but at the level of
accuracy of our calculations below, this contribution will not be
significant (the calculation is currently in progress by one of
us).
The effective potential for $\phi$ is then
\beq
V_{eff}(\phi) = a \alpha_s(T)^2 T^4 \ln(\phi^2),
\eeq
which is obtained from the formula for the
free energy as a function of temperature
where
\beq
a = {3 N_g \over 288} (5N_f/4 + 7N/2)
\eeq
is a bit larger than one.

\section{Estimating the Baryon Number}

We will now study the effects of these thermal terms for various
values of $n$.  The full parameter space we might explore is very
large.  Possible parameters include:  the coefficients of the
$\phi^n$ terms in the superpotential, the coefficients of the $A$
terms proportional to $H$, the coefficient of the scalar mass
terms proportional to $H^2$, the size of CP violating couplings,
the coefficients of the curvature independent mass and $A$ terms,
the value of $m_{3/2}$, the reheat temperature, $T_R$,
the value of the Hubble constant during inflation, $H_I$,
and so on.  In addition, there are many
discrete choices we might examine, including the particular flat
direction.  We will leave a more careful survey of the parameter
space to subsequent work\cite{anisimov}, and here
instead will choose some particular points in the parameter space
in order to illustrate the possibilities.  We will see that for
$n=1$ directions, with small Yukawa couplings (less that about
$10^{-2}$) that it may just barely be possible to produce an
adequate asymmetry.  For $n=2$, the results are quite sensitive to
different parameters.  In particular, depending on parameters,
different types of thermal effects are important.  For $n=3$ and
$n=4$, it is not difficult to produce an appreciable asymmetry,
but thermal effects do alter the asymmetry relative to the
expectations in \cite{drt}, in some cases by several orders of magnitude.

Some words about notation are in order.  We will, in any given case, refer to
$H_o$ as the value of the Hubble constant when the system begins
to oscillate.  $H_I$ will be the value of the Hubble constant
during inflation, which we will generally take to be $10^{13}$
GeV.  We will usually take the reheating temperature to be
$T_R= 10^{10}$ GeV.  All of our estimates are easily modified
for other choices of these parameters.

Let us consider, first, the case where some of the fields which
couple to the condensate are in
thermal equilibrium.
As a model, we take:
\beq
V= (-H^2 + m_{3/2}^2 + y^2 T^2)\vert \phi \vert^2 + \vert
{\partial W \over \partial \phi} \vert^2  + a H W + b{H^2 \over M}
W + A m_{3/2} W ~~~~~~W = {\phi^{n+3} \over M^n}.
\eeq
We can at first ignore the $m_{3/2}$ terms.  Then we can
ask:  under what circumstances are the fields in thermal
equilibrium.  We have seen that in the case $n=1$, the
fields can easily be in equilibrium immediately after
inflation.  Consider the case $n=2$.  Again taking units
with $10^{13}$ GeV $=1$
for the Hubble constant during inflation and for the value
of the initial temperature  we have $\phi^6 \sim H^2 M^4$.
Requiring $y \phi \sim T$, with $T \sim H^{1/4}$, gives
\beq
H \sim y^{-12} M^{-8}.
\eeq
Obviously the result is quite sensitive to $y$ and to constants of
order one (e.g. the Planck mass vs. the reduced Planck mass in
theses formulas); for $y=10^{-3}$ and $M=10^{5}$, one obtains
$H_o=10^{-4}= 10^9$ GeV.  In this case, $y \phi = 10^{-1} \approx
T$ at this time,
so thermal effects are important.
If one used the Planck mass this would
become $H = 10^{-12} = 10$ GeV, so the $m_{3/2}$ term
is already dominant.  (When oscillations begin, $y \phi \approx
1$ while $T \sim 10^{-2.5}$, so thermal effects are irrelevant).

Now we estimate the baryon number.  In the case
$n=1,2$, we expect that the $A$ term suppressed by $I /M$ discussed
in equation \ref{isquared} to be the principle source of baryon
number.  So we expect the rate of change of the baryon number
per unit time is given roughly by: \beq {d n_B \over dt} = H
{\phi^4 \over M} {b I \over M} \sin(\delta) \eeq
where $\sin(\delta)$ represents an appropriate combination of
$CP$-violating phases.
If $H_o$ and
$\phi_o$ are the values of $\phi$ and $H$ when oscillations begin,
then the baryon number is of order:
\beq n_B = {\phi_o^4
\over M} b {H_o \over H_I} \sin(\delta). \eeq
To estimate the
final baryon/entropy ratio we can multiply by $t^2 = H_o^{-2}$, and
divide by $T_R^3 t_D^2$, where $T_R$ is the reheating temperature
and $t_D \sim {m_I^3 \over M^2}$ is the decay time. Putting this together, for $T_R=
(H M_I^3)^{1/4}$, we obtain the estimate, for our canonical inflaton
model:
\beq {n_B \over T^3} = b \sin(\delta)
 {\phi_o^4 \over M H_o
T_R^3 t_d^2}. \eeq

In the case $n=1$, $\phi_o^2 \approx
H_o M$, and $H_o \approx y^{4/3} M_I$, so plugging in the formulas above,
gives the crude estimates:
\beq
n=1:  n_B  = b  ({y \over 10^{-2}})^{4/3}  \sin(\delta)
10^{-14}.
\eeq
This is four order of magnitudes less then desired ratio.
Actual numerical study of
the differential equation for the evolution of the scalar field
gives results, for a range of parameters, within an order of magnitude
of this. However, if the reheating temperature is somewhat lower one might be
able to produce larger assymetry.

The case $n=2$ (with the choice $M=10^{18}$ discussed
above) is more promising.  Now $\phi^6 = M^4 H^2$.
Oscillations begin much later, as indicated above.  Taking the
case where oscillations start at $H_o= 10^7$ GeV,
we have for the baryon density, proceeding as before,
\beq
{n_B t^2 \over T^3} = {M^{1/3} H^{2/3}_o \over
T_R^3 t_D^2}b \sin(\delta) \approx  (10^{-8}-10^{-10})~b \sin(\delta)
\eeq
As before, if the reheat temeperature is lower one can produce
an even larger asymmetry.

For $n=3$, the field typically comes to equilibrium only after the
inflaton has decayed.\\

In the case $n \ge 3$ (or $n=2$ or $n=1$ if the Yukawa coupling is not too
small), we should ask about the effects of the
thermal potential.
In such cases the potential we wish to study has the form:
\beq
V= (-H^2 + m_{3/2}^2 )\vert \phi \vert^2+ a T^4 \ln(\vert \phi
\vert^2) + \vert
{\partial W \over \partial \phi} \vert^2  + a H W + b{H^2 \over M}
W + A m_{3/2} W ~~~~~~W = {\phi^{n+3} \over M^n}.
\eeq
We first should ask:  when is the $T^4$ term relevant.  Comparing
the derivatives of the $H^2 \phi^2$ and the $T^4$ terms gives:
\beq
n=2:  H_o = 10^{-2} M^{-4/5} \approx 10^{-6}~~~~~~
n=3: H_o= 10^{-2} M^{-1} = 10^{-7}~~~~~~~
n=4:  H_o = 10^{-9}.
\eeq
This is before the decay of the inflaton in each case.  At this
time, for $n=3$, $\phi \sim 2 \times 10^{15}$ GeV, while $T \sim 10^{11}$,
so thermal effects should be unimportant, except for extremely
small Yukawa couplings.  Evaporation typically occurs well after
the decay of the inflaton.
To estimate the decay time, as in \cite{drt}, one notes that the
leading couplings of thermal particles to the condensate are
through dimension $5$ operators, so that the annihilation rate
behaves as:
\beq
\Gamma_{ann} \approx ({\alpha \over 4\pi})^3 {T^3 \over \phi^2}
\eeq
where $\alpha$ is an appropriate gauge coupling (two factors
of $\alpha$ arise from loops).  This is typically less than $H$
until the inflaton decays.  After this time, as in \cite{drt},
the rate slows more rapidly than $H$, and evaporation eventually
occurs.

Again, we
can make a crude estimate of the baryon number.
For the case $n=2$, $\phi^6 = M^4 H^2$; for $n=3$, $\phi^8 = M^6 H^2$.
We expect that the baryon
number is suppressed by $m_{3/2} \over H_o$, the ratio of the
curvature independent to the curvature-dependent $A$ terms, so we
estimate:
\beq
n=2:  n_B t^2 \approx {\phi_o^5 \over M^2 H^2}{m_{3/2} \over H_o}
\sin(\delta) \approx 10^{18} \sin(\delta) {\rm GeV}
\eeq
\beq
n=3:
n_B t^2 \approx {\phi_o^6 \over M^3 H_o^2}{m_{3/2} \over
H_o} \sin(\delta) \approx 10^{23} \sin(\delta)
 {\rm GeV}
\eeq
where, as usual, $\sin(\delta)$ represents some combination of
CP-violating phases.  Thus the baryon to photon ratio is of order
$10^{-10}
\sin(\delta)$ in the $n=2$ case, which is clearly in an
interesting range.  For the case $n=3$, the ratio is roughly
$10^{-4} \sin(\delta)$, which can be substantial even for small
values of the CP violating phases.

It is interesting to compare this result to what is obtained
without the inclusion of thermal effects, when oscillation starts
at $H\sim m_{3/2}$, i.e. about four (n=2) or two (n=3) orders of magnitude
later.  The present baryon to
photon ratio is about six orders of magnitude
smaller in the first case, three orders of magnitude smaller in the
second.  This
difference traces to two factors:  first, there is the suppression
by $m_{3/2}/H_o$; second, the baryon number violating term times
$t^2$ scales as $H_o^{-1/2}$, accounting for another order of
magnitude.

For the case $n=4$ the results are similar.  One still produces a
substantial baryon number, now suppressed by about two orders of
magnitude relative to the non-thermal analysis.

Numerical study, however, indicates that these estimates are not
always reliable.  The baryon number is often significantly larger.
A systematic study of the parameter space will appear
elsewhere\cite{anisimov}.  For the moment, our main point is that
the inclusion of thermal effects, even for rather flat potentials,
can significantly alter the prediction of the asymmetry.

\section{Q Balls}

In the last few years, several authors have observed that
supersymmetric theories often contain $Q$-balls in their spectra,
and that these might be produced in the evolution of the
Affleck-Dine condensate\cite{qballs}.  The usual picture of
$Q$-ball formation is to note that under certain circumstances,
the evolution of the homogeneous condensate is unstable for small
momenta, and to argue that this instability is likely to lead to
$Q$-ball formation.  At this point, there is support for such a
picture from numerical simulations in some cases\cite{simulations}.

Before considering thermal effects,
we would comment that it is possible for a condensate to
exhibit instability even when it carries no baryon number, so it
is by no means clear that any instability one finds is a signal of
$Q$-ball formation.  For example, the potentials associated with
gauge mediated models have such instabilities, whether or not
the scalar fields carry non-trivial phases.
More detailed study is then necessary to determine whether
$Q$-balls form.

More relevant to
our present discussion, though, is the fact that the evolution of
the condensate, in light of the various thermal effects described
here, is rather different than usually expected.  This point has
already been noted in ref. \cite{campbell}.  The issue, in
general, is whether $Q$-balls can form before the evaporation of
the condensate, and whether they survive
the evaporation process.

As we have seen, for a range
of $n$, thermal effects control the evolution of the condensate.
In the case where the $\chi$ field is in equilibrium, the potential
includes not only quadratic terms, but a negative, cubic term,
\beq
V(T) = ag^2 T^2 \vert \phi \vert^2 - b g^3 T \vert \phi \vert^3
\eeq
It is easy to check that this equation satisfies the conditions of
\cite{qballs} for growing instabilities.  On the other hand, it is
also true that any $Q$-ball which forms will evaporate in much the
same way as the condensate.  We will present a more detailed
analysis of this problem in a subsequent publication (including
issues such as the evolution of the instabilities in the thermal
potential).

In cases of larger $n$, where the $\ln(\phi^2)$ terms in the
potential are important, evaporation eventually destroys
the condensate but not necessarily the $Q$-balls.  Here the required
estimates are similar to those performed in \cite{drt}.  There are
now particles in thermal equilibrium which couple to the
condensate through higher dimension operators.  If the operators
are dimension five, the interaction rate is of order
\beq
\Gamma = {\alpha_s^3 \over 16 \pi^2 \phi^2} T^3
\eeq
In the case $n=2$ (assuming that there are no
fields in equilibrium with the condensate), for example, one finds that the condensate
eventually evaporates through interactions with the thermal bath.
Consider, first, the homogeneous condensate.  If one substitutes
the expressions for $T$ and $\phi$ as a function of $H$ in the expression
for the reaction rate, one finds that immediately after inflation,
\beq
{\Gamma \over H} \approx 10^{-8}
\eeq
and that it grows as $H^{-5/6}$.   So interactions with the
condensate are expected to destroy the condensate shortly after crossover.
As for Q-balls, on the one hand, the mean free paths of the individual $\chi$
particles is large compared to the size of the $Q$-balls.
On the other, the scattering off the particles
in any would-be Q-balls is very
rapid.

In the case $n \ge 3$, the condensate evaporates much
later, and Q-balls are likely to survive.
More detailed investigation
and further simulations would be worthwhile in these cases.

\section{Conclusions}

Thermal effects significantly alter the predictions for
baryogenesis due to the coherent evolution of scalar fields.
In agreement with \cite{campbell}, in the case that the potential
is not terribly flat (e.g. $n=1$) and that the Yukawa
couplings are small enough, we have seen that the thermal
potential plays an important role in the evolution of the
condensate.  Oscillations start earlier than in the picture of
\cite{drt}, and the baryon number is produced earlier.  This leads
to a more promising result than predicted by \cite{drt}, who
argued that the rapid evaporation of the condensate in these cases
could prevent any asymmetry from developing.  On the other hand,
our estimates are less optimistic than those of \cite{campbell} in
these cases, since, as we have argued, $CP$ violation is
suppressed.

We have seen that thermal effects are potentially important even in cases in
which the potential is relatively flat ($n=2,3,4$).  After
integrating out heavy particles, interactions with the remaining
light particles, while suppressed, can be significant.  This can
alter by several orders of magnitude the predictions for the
asymmetry.  We have also seen that in some cases, the evaporation
of the condensate tends to eliminate also any would-be $Q$-balls.
The point is that the light particles in the plasma have a long
mean free path that the Q-balls are transparent,
but also have large enough interactions with
the condensate to destroy it.

For large enough $n$, thermal considerations are irrelevant.
It is certainly of interest to further explore the possible roles
of $Q$-balls in these cases.  More refined numerical studies, both
of the condensate evolution described here, and of the non-linear
evolution of fluctuations, will be necessary to develop a
comprehensive picture.

\noindent
{\bf Acknowledgements:}

\noindent
This work supported in part by the U.S.
Department of Energy.  M.D. wishes to thank Scott Thomas for
discussions and comments on the manuscript.


\end{document}